\documentclass[aps,pra,reprint,groupedaddress,showpacs]{revtex4-1}
\usepackage{graphicx}
\usepackage{latexsym,amssymb,amsmath}
\usepackage{bm}
\usepackage[caption=false]{subfig}
\usepackage{xcolor}
\usepackage{mathrsfs}
\usepackage[colorlinks=true,
urlcolor=blue,
linkcolor=blue,
citecolor=blue
]{hyperref}

\begin{document}
\title{Wideband tunable infrared topological plasmon polaritons in dimerized chains of doped-silicon nanoparticles}

\author{B. X. Wang}
\author{C. Y. Zhao}
\email{changying.zhao@sjtu.edu.cn}

\affiliation{Institute of Engineering Thermophysics, School of Mechanical Engineering, Shanghai Jiao Tong University, Shanghai, 200240, China}
\date{\today}T

\begin{abstract}
We investigate the topological plasmon polaritons (TPPs) in one-dimensional dimerized doped silicon nanoparticle chains, as an analogy of the topological edge states in the Su-Schrieffer-Heeger (SSH) model. The photonic band structures are analytically calculated by taking all near-field and far-field dipole-dipole interactions into account. For longitudinal modes, it is demonstrated that the band topology can be well characterized by the complex Zak phase irrespective of the lattice constant and doping concentration. By numerically solving the eigenmodes of a finite system, it is found that a dimerized chain with a nonzero complex Zak phase supports nontrivial topological eigenmodes localized over both edges. Moreover, by changing the doping concentration of Si, it is possible to tune the resonance frequency of the TPPs from far-infrared to near-infrared, and the localization length of the edge modes are also modulated accordingly. Since these TPPs are highly protected modes that can achieve a strong confinement of electromagnetic waves and are also immune to impurities and disorder, they can provide a potentially tunable tool for robust and enhanced light-matter interactions light-matter interaction in the infrared spectrum.

\end{abstract}

\maketitle
\section{Introduction}
The advent and rapid development of topological photonics provide great opportunities to study the physics of topological phases of matter in the optical context \cite{ozawa2018topological,riderJAP2019,xieOE2018}. A key feature of topological photonic systems is that they can hold topologically protected modes of light, which are highly localized over the system boundaries and can propagate without any backscattering processes even in the presence of disorder and impurities \cite{luNPhoton2014,khanikaevNPhoton2017,ozawa2018topological}. These topological modes can be utilized to achieve precise, robust and local control of light, which thus facilitate novel photonic devices such as unidirectional waveguides \cite{poliNComms2015}, optical isolators \cite{el-GanainyOL2015} and topological lasers \cite{stjeanNaturephoton2017,partoPRL2018,zhaoNaturecomms2018}, and also open pathways for robust quantum information and quantum computation by creating topologically protection of multiphoton states and quantum entanglement \cite{blanco-redondoScience2018,wangOptica2019,wang2019topologically,michelleNanophoton2019}. 

Among the topological photonic systems, recently, there has been a growing interest in topological plasmon polaritons (TPPs) \cite{lingOE2015,downingPRB2017,pocockArxiv2017,downing2018topological,pocockNanophoton2019}. One of the simplest platforms that can realize topological plasmon polaritons derives from the Su-Schrieffer-Heeger (SSH) model \cite{suPRL1979}, which is based on one-dimensional (1D) periodic, dimerized plasmonic nanoparticle (NP) chains \cite{lingOE2015,downingPRB2017,pocockArxiv2017,downing2018topological,pocockNanophoton2019}. Other topological plasmon polaritons in 1D nanostructures are realized in plasmonic waveguide arrays \cite{chengLPR2015}, multilayered graphene systems \cite{geOE2015,xuAppsci2019}, graphene nanoribbon arrays \cite{zhangPRB2018}, metagate-defined 1D graphene metasurfaces encapsulated by two hBN layers \cite{fanNanophoton2019} and so on. Topological Majorana plasmon polaritons were also realized in a quasi-1D (zig-zag) plasmonic nanodisk chain by mimicking the Kitaev model \cite{poddubnyACSPhoton2014}. Moreover, it was shown that topological Tamm plasmon polaritons \cite{wangOE2018} can be achieved with reduced loss by using periodic metal-insulator-metal waveguides. In higher dimensions, topological protected plasmonic modes have been realized in graphene superlattices \cite{panNaturecomms2017} and metallic ring resonator arrays \cite{gaoNaturecomms2016}. It was also demonstrated that two-dimensional nanoparticle or nanowire lattices can be engineered to support TPPs \cite{fernique2019plasmons,mengOQE2019}. For example, by mimicking the quantum spin Hall effect in topological insulators, pseudospin dependent edge states were realized in 2D plasmonic metasurfaces \cite{proctor2019exciting}. Note another approach to topological plasmon polaritons is to couple light into the topologically protected plasmons in some topological insulators to obtain topological plasmon polaritons, rather than by engineering the nanostructures \cite{iorsh2019plasmon}. As a result of the unique combination of topological protection and nanoscale light confinement arising from plasmonic excitations, these TPPs are very promising for achieving robust and deep-subwavelength scale light-matter interactions. For instance, the modal wavelength of topologically bounded plasmonic modes in multilayered graphene systems can be squeezed as small as 1/70 of the incident wavelength \cite{xuAppsci2019}. A recent theoretical study showed that low-power-consumption and highly-integrated four-wave mixing processes can be observed through the TPP modes in graphene metasurfaces \cite{you2019four-wave}. 



In this paper, we aim to realize and investigate topologically protected plasmon polaritons in 1D dimerized doped silicon (Si) NP chains as an extension of the SSH model. The choice of doped Si is stimulated by the motivation of achieving tunable TPPs that can be potentially modulated to emerge in a relatively wide spectral range. By now, most works on TPPs based on the optical analog of SSH model using metallic nanoparticles have been focused on some specific wavelength ranges  \cite{lingOE2015,downingPRB2017,downing2018topological,pocockArxiv2017,pocockNanophoton2019} that are relatively difficult to tune. Although TPPs have been theoretically shown to be largely tunable using electrical gating and carrier doping methods in graphene-based nanostructures \cite{geOE2015,zhangPRB2018,xuAppsci2019,fanNanophoton2019}, the experimental demonstration still remains to be difficult. In addition, our previous studies on topological optical modes in cold atom lattices \cite{wang2018topological} and silicon carbide NP arrays \cite{wangPRB2018b} are, to some extent, also limited to narrow wavelength ranges that are determined by the innate excitations (atomic levels and longitudinal phonons, respectively) in the systems under investigation. In this circumstance, heavily doped Si that supports surface plasmon resonances has a great potential in achieving topological optical excitations within a comparably wide spectral range in a relatively easy-to-access way \cite{basuJHT2010,ginnJAP2011,liuJHT2013,gorguluJO2016,salmanADOM2017,tervoAPL2019}. Another motivation is that it is possible to tune these topological plasmon polaritons in a way such that they can be thermally excited at in a wide range of working temperatures, which may contribute to a large enhancement to near-field heat transfer and near-field thermophotovoltaics \cite{watjenAPL2016,fernandezPRL2017,limJQSRT2018,desutterNaturenano2019}.

More precisely, in this paper, small doped Si NPs are treated as electric dipoles and the photonic band structures of the dimerized 1D chains are analytically calculated by taking all near-field and far-field dipole-dipole interactions into account based on the coupled-dipole model. For longitudinal modes, it is demonstrated that the band topology can be well characterized by the complex Zak phase, which indicates a topological phase transition when the dimerization parameter changes from less than 0.5 to larger than 0.5, irrespective of the lattice constant and doping level. By numerically solving the eigenmodes of a finite system as well as their inverse participation ratios (IPRs), a dimerized chain with a nontrivial (nonzero) complex Zak phase is found to be able to support nontrivial topological eigenmodes localized over both of its edges. Moreover, by changing the doping type and concentration of doped Si, it is possible for us to tune the resonance frequency of the TPPs from far-infrared to near-infrared, potentially via methods like optical pumping or electrical gating \cite{liuJHT2013,ferreraJOSAB2017,tervoAPL2019}, and the localization lengths of the edge modes are also modulated accordingly. As a result, these highly protected TPP modes provide an efficient tool for enhancing light-matter interaction in a relatively wide spectral band in the infrared in a tunable fashion.


\section{Model}\label{model}
In this section, we briefly describe the analytical and numerical models used in this study. The model is based on classical electrodynamics and no nonlinear and quantum effects are taken into account. We first provide the analytical expressions for the permittivity of doped silicon, as provided in Ref.\cite{basuJHT2010}. Then we describe the electric dipole approximation and the coupled-dipole model. On this basis, the analytical calculation of the Bloch band structure for an infinitely long chain and the eigenmode distribution for a finite lattice are given \cite{wangPRB2018b}. 
\subsection{Model of permittivity of doped silicon}
The permittivity function of doped Si can be modeled by a Drude model as \cite{basuJHT2010,fernandezPRL2017}
\begin{equation}\label{permittivity}  \varepsilon_p(\omega)=\varepsilon_\infty-\frac{\omega_p^2}{\omega^2+i\gamma\omega},
\end{equation}
where $\omega$ is the angular frequency of the driving field in the unit of $\mathrm{cm}^{-1}$, $\varepsilon_\infty=11.7$ is the high-frequency limit of the permittivity, $\omega_p$ is the plasma frequency, and $\gamma$ is the damping coefficient, which both depend on the doping type and concentration.
Here we adopt the empirical formulas from Ref.\cite{basuJHT2010}. The parameters in this model are related to the effective mass and mobility as $\omega_p=\sqrt{Ne^2/m^*\varepsilon_0}$ and $\gamma=e/m^*\mu$, where $N$ is the carrier concentration at the given doping concentration, $m^*$ is the effective mass of carries, $\mu$ is the mobility, $e$ is the electric charge of an electron and $\varepsilon_0$ is the permittivity of the vacuum. In heavily doped silicon, the effective mass of electron or hole is assumed to be independent of the carrier concentration and frequency, which is taken as $0.27m_0$ and $0.37m_0$ for electron and hole, respectively, where $m_0$ is the mass of a free electron in vacuum. At a given doping concentration, the mobility of $n$-type doped Si is \cite{basuJHT2010}
\begin{equation}
\mu = \mu _ { 1 } + \frac { \mu _ { \max } - \mu _ { 1 } } { 1 + \left( N _ { e } / C _ { r } \right) ^ { \alpha } } - \frac { \mu _ { 2 } } { 1 + \left( C _ { s } / N _ { e } \right) ^ { \beta } },
\end{equation}
where $\mu_1=68.5~\mathrm{cm^2/V s}$, $\mu_\mathrm{max}=1414 ~\mathrm{cm^2/V s}$, $\mu_2=56.1~\mathrm{cm^2/V s}$, $C_r=9.2\times 10^{16}~\mathrm{cm}^{-3}$, $C_s=3.41\times 10^{20} ~\mathrm{cm}^{-3}$, $\alpha=0.711$, $\beta=1.98$, and $N_e$ is the electron concentration. And the mobility of $p$-type doped Si is 
\begin{equation}
\mu = \mu _ { 1 } \exp \left( - p _ { c } / N _ { h } \right) + \frac { \mu _ { \max } } { 1 + \left( N _ { h } / C _ { r } \right) ^ { \alpha } } - \frac { \mu _ { 2 } } { 1 + \left( C _ { s } / N _ { h } \right) ^ { \beta } },
\end{equation}
where $\mu_1=44.9~\mathrm{cm^2/V s}$, $\mu_\mathrm{max}=470.5 ~\mathrm{cm^2/V s}$, $\mu_2=29.0~\mathrm{cm^2/V s}$, $C_r=2.23\times 10^{17}~\mathrm{cm}^{-3}$, $C_s=6.10\times 10^{20} ~\mathrm{cm}^{-3}$, $\alpha=0.719$, $\beta=2$, $p_c=9.23\times 10^{16}~\mathrm{cm}^{-3}$, and $N_e$ is the electron concentration.

The carrier concentration is determined by the degree of ionization $\zeta$, which can be modeled empirically as \cite{kuzmiczSSE1986}
\begin{equation}
\zeta = 1 - A \exp \left\{ - \left[ B \ln \left( N_d / N _ { 0 } \right) \right] ^ { 2 } \right\},
\end{equation}
where $N_d$ is the doping concentration, and the constants $A$, $B$, and $N_0$ are determined as follows. For $n$-type Si: $A=0.0824\Theta^{-1.622}$, $N_0=1.6\times10^{18}\Theta^{0.7267}$, and $B=0.4722\Theta^{0.0652}$ for $N_d<N_0$, and $B=1.23-0.3162\Theta$ for $N_d\geq N_0$, and $\Theta=T/300$ is a reduced temperature. For $p$-type Si: $A=0.2364\Theta^{-1.474}$, $N_0=1.577\times10^{18}\Theta^{0.46}$, and $B=0.433\Theta^{0.2213}$ for $N_d<N_0$, and $B=1.268-0.338\Theta$ for $N_d\geq N_0$. 

With these parameters, above equations for the permittivity of doped Si were shown to agree well with experimental measurements of the transmittance and reflectance of doped Si wafers up to a doping concentration of $N_d=10^{21}~\mathrm{cm}^{-3}$ in the wavelength region from $2\mathrm{\mu m}$ to $20 \mathrm{\mu m}$ at room temperature.

\subsection{The electric dipole approximation and coupled-dipole model}
\begin{figure}[htbp]
	\centering
	\subfloat{
		\includegraphics[width=0.8\linewidth]{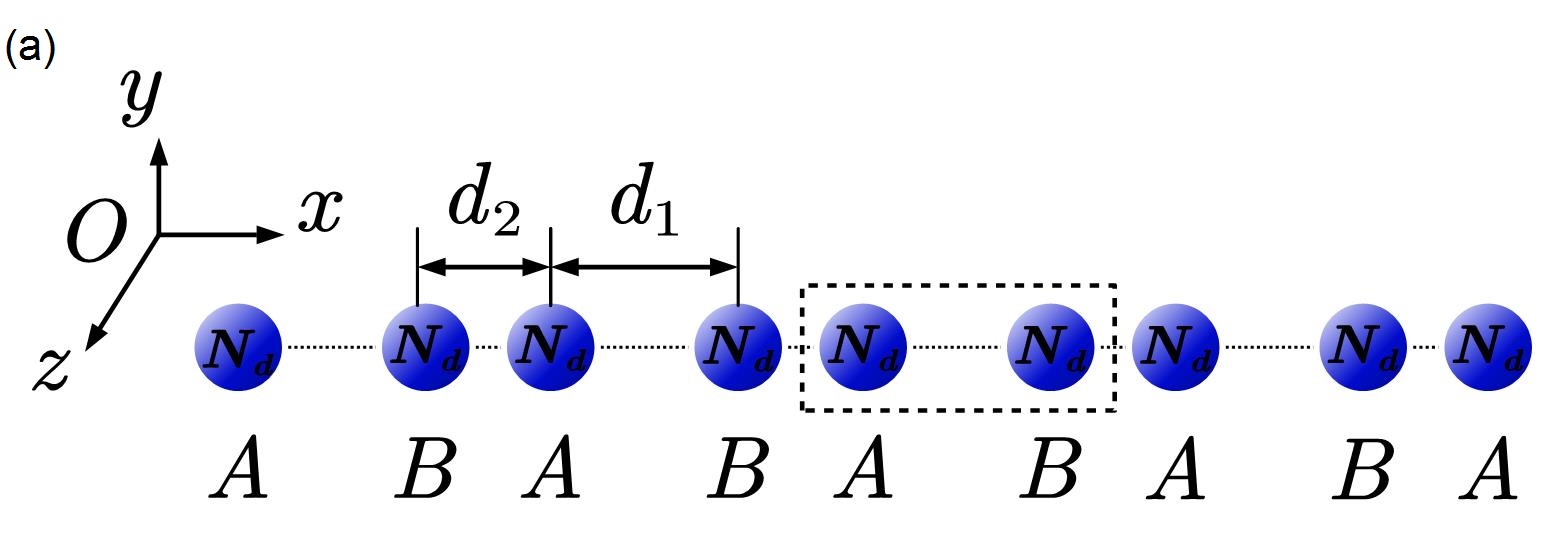}\label{schematic_dSi}
	}
	\hspace{0.01in}
	\subfloat{
		\includegraphics[width=0.48\linewidth]{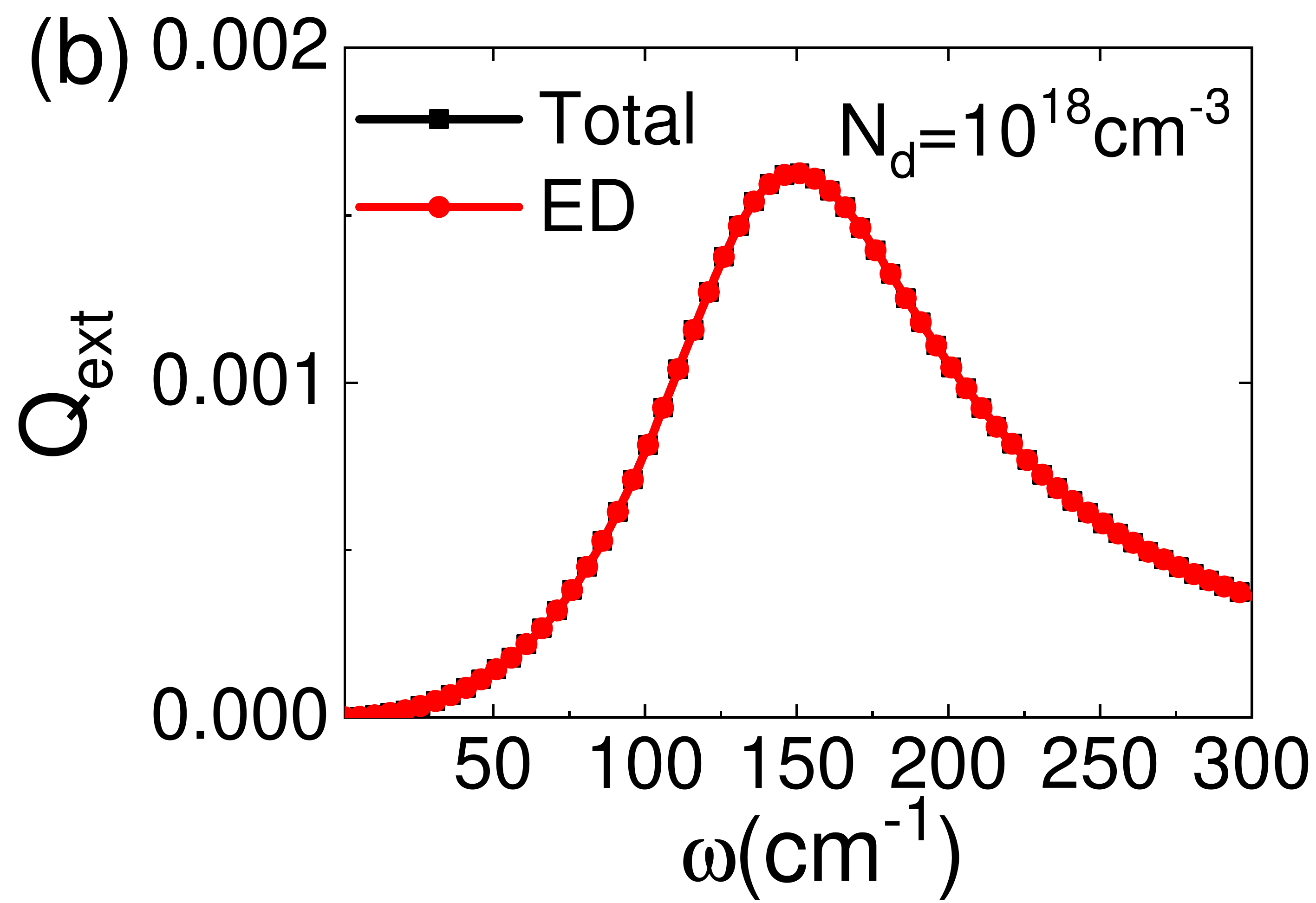}\label{qextsingle}
	}
	\hspace{0.01in}
   \subfloat{
	\includegraphics[width=0.45\linewidth]{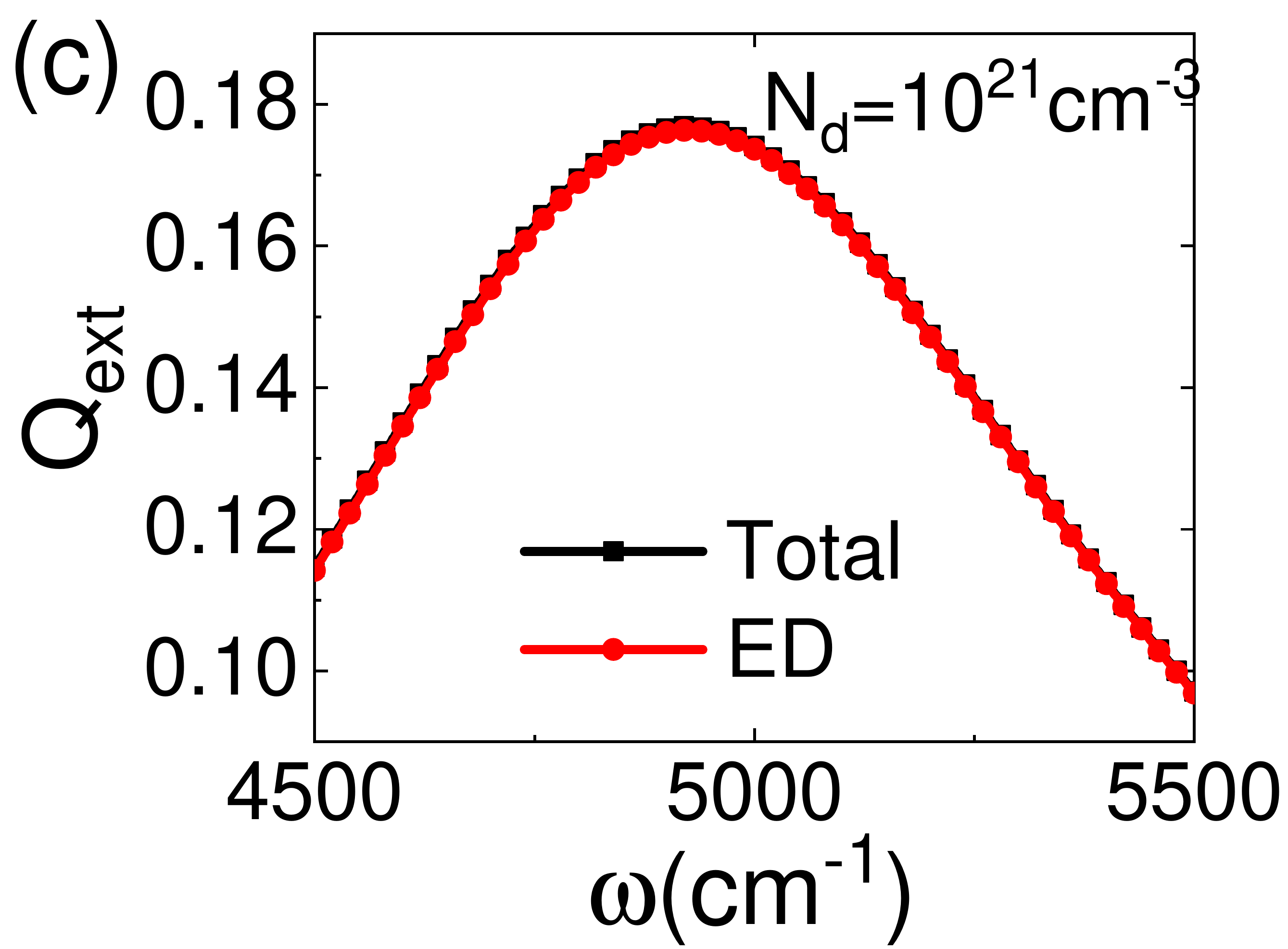}\label{qextsingle2}
  }
	\caption{(a) Schematic of the dimerized doped Si nanoparticle chain. The nanoparticles are identical with a doping concentration of $N_d$ and each sublattice with different lattice constants are denoted by $A$ and $B$ respectively. The inequivalent inter-particle spacings are represented by $d_1$ and $d_2$. The unit cell is denoted by the dashed rectangle. (b,c) Extinction efficiency $Q_\mathrm{ext}$ of a single $n$-doped Si NP under different doping concentrations, with a radius of $a=0.1\mathrm{\mu m}$. A comparison is made between the Mie theory, compared and the electric dipole (ED) approximation.}\label{schematicandmie}
	
\end{figure} 

We consider a 1D dimerized chain composed of spherical doped Si NPs schematically shown in Fig.\ref{schematic_dSi}. The NP chain is well aligned along the $x$-axis, where the dimerization is introduced by using inequivalent spacings $d_1$ and $d_2$ for the two sublattices, denoted by $A$ and $B$. Here we define the dimerization parameter as $\beta=d_1/d$ where $d=d_1+d_2$ is the overall lattice constant.  Such dimerization leads to different “hopping” amplitudes of photons in different directions, mimicking
the SSH model for electron hopping \cite{suPRL1979,asboth2016short}. Note in the presence of near-field and far-field dipole-dipole interactions, the physical picture is more complicated than nearest-neighbor and Hermitian
hopping amplitudes assumed in the conventional SSH model \cite{asboth2016short,wangPRB2018b}.

Without loss of generality, herein we set the radius of the spherical doped Si NP as $a=0.1\mathrm{\mu m}$, which is much smaller than the wavelength of interest. In this situation, the single particle extinction efficiency $Q_\mathrm{ext}=C_\mathrm{ext}/(\pi a^2)$ is calculated using the Mie theory as shown in Fig.\ref{qextsingle}, where $C_\mathrm{ext}$ is the extinction cross section. Actually, such a small NP can be well modeled by the electric dipole (ED) approximation. By considering electric dipole excitations only, the EM response of an individual doped Si NP is described by the dipole polarizability with the radiative correction, which is given as \cite{tervoPRMater2018,markelPRB2007,parkPRB2004}
\begin{equation}\label{radiativecorrection}
\alpha(\omega)=\frac{4\pi a^3\alpha_0}{1-2i\alpha_0(ka)^3/3},
\end{equation}
where 
\begin{equation}
\alpha_0(\omega)=\frac{\varepsilon_p(\omega)-1}{\varepsilon_p(\omega)+2}.
\end{equation}

The extinction efficiency calculated under the ED approximation using the polarizability with the radiative correction under two different doping concentrations at a temperature of 300 K are shown in Figs.\ref{qextsingle} and \ref{qextsingle2}, and a good agreement with the exact Mie theory is observed. We can also note that different doping concentrations lead to localized surface plasmon resonances (LSPR) with different resonance frequencies in the doped Si NPs, ranging from the far-infrared to near-infrared for $N_d=10^{18}~\mathrm{cm}^{-3}$ to $N_d=10^{21}~\mathrm{cm}^{-3}$, which provide the basis of the potentially tunable collective SPP excitations like TPPs.

For this 1D NP chain, when the distance between the centers of different spherical NPs is less than $3a$, the electromagnetic interactions are exactly captured by the coupled-dipole equations \cite{parkPRB2004,markelPRB2007,tervoPRMater2018}:
\begin{equation}\label{coupled_dipole_eq}
\mathbf{p}_j(\omega)=\alpha(\omega)\left[\mathbf{E}_\mathrm{inc}(\mathbf{r}_j)+\frac{\omega^2}{c^2}\sum_{i=1,i\neq j}^{\infty}\mathbf{G}_0(\omega,\mathbf{r}_j,\mathbf{r}_i)\mathbf{p}_i(\omega)\right],
\end{equation}
where $c$ is the speed of light in vacuum. $\mathbf{E}_\mathrm{inc}(\mathbf{r})$ is the external incident field and $\mathbf{p}_j(\omega)$ is the excited electric dipole moment of the $j$-th NP. $\mathbf{G}_{0}(\omega,\mathbf{r}_j,\mathbf{r}_i)$ is the free-space dyadic Green's function describing the propagation of field emitting from the $i$-th NP to $j$-th NP \cite{markelPRB2007}. This model takes all types of near-field and far-field dipole-dipole interactions into account and is thus beyond the traditional nearest-neighbor approximation, which is implemented in the SSH model \cite{suPRL1979,lingOE2015}. 


\subsection{Infinite chains}\label{model_infinite}
For 1D chains, according to the polarization direction of the dipole moments of the NPs, the electromagnetic eigenmodes can be divided into transverse and longitudinal modes \cite{weberPRB2004}. In our system, the dipole moments of the NPs in the longitudinal eigenmodes are polarized along the $x$-axis, while those in the transverse eigenmodes are polarized perpendicular to the $x$-axis. In this paper, we are mainly concerned with the longitudinal eigenmodes because transverse ones are more strongly coupled to the free-space radiation and the band gap of transverse eigenmodes is much narrower, which makes it difficult to observe these topological eigenmodes experimentally \cite{wang2018topological,wangPRB2018b}.

For an infinitely periodic chain, by applying the Bloch theorem, the wavefunction (or dipole moment distribution) of the longitudinal Bloch eigenmode with a momentum $k_x$ along the $x$-axis can be expressed as $p_{m_i,k_x}(\omega)\exp{(ik_xx_i)}$ with $m_i=A, B$, and inserting this expression into Eq.(\ref{coupled_dipole_eq}) with zero incident field leads to
\begin{equation}\label{coupled-dipole_bloch}
\begin{split}
&\frac{\omega^2}{c^2}\sum_{i=1,i\neq j}^{N}G_{0,xx}(\omega,\mathbf{r}_j,\mathbf{r}_i)p_{m_i,k_x}(\omega)\exp{(ik_xx_i)}\\&=\alpha^{-1}(\omega)p_{m_j,k_x}(\omega)\exp{(ik_xx_j)}.
\end{split}
\end{equation}
Here the $xx$-component of the Green's function is used:
\begin{equation}\label{Gxx}
G_{0,xx}(x)=-2\Big[\frac{i}{k|x|}-\frac{1}{(k|x|)^2}\Big]\frac{\exp{(ik|x|)}}{4\pi |x|},
\end{equation}
where $k=\omega/c$ is the free-space wavenumber. More specifically, by explicitly carrying out the summations, Eq.(\ref{coupled-dipole_bloch}) is rewritten in the following form
\begin{equation}\label{eigenvalue_long}
\frac{\omega^3}{c^3}\left(\begin{matrix}
a_{11}(k_x) & a_{12}(k_x)\\
a_{21}(k_x) & a_{22}(k_x)
\end{matrix}\right)\left(\begin{matrix}p_{A,k_x}\\p_{B,k_x}\end{matrix}\right)=\alpha^{-1}(\omega)\left(\begin{matrix}p_{A,k_x}\\p_{B,k_x}\end{matrix}\right),
\end{equation}
where the diagonal matrix elements are evaluated as \cite{wangPRB2018b}
\begin{equation}
\begin{split}
&a_{11}(k_x)=a_{22}(k_x)=-i\frac{\mathrm{Li}_2(z^+)+\mathrm{Li}_2(z^-)}{2\pi k^2d^2}\\&+\frac{\mathrm{Li}_3(z^+)+\mathrm{Li}_3(z^-)}{2\pi k^3d^3}.
\end{split}
\end{equation}
Here $z^+=\exp{(i(k+k_x)d)}$ and $z^-=\exp{(i(k-k_x)d)}$, and $\mathrm{Li}_s(z)$ is polylogrithm (or Jonqui\'ere's function) $\mathrm{Li}_s(z)=\sum_{n=1}^\infty z^n/n^s$ sum series \cite{NISThandbook}. The off-diagonal elements are \cite{wangPRB2018b}
\begin{equation}\label{a12Leq}
\begin{split}
&a_{12}(k_x)=\Big[-i\frac{\Phi(z^+,2,\beta)}{2\pi k^2d^2}+\frac{\Phi(z^+,3,\beta)}{2\pi k^3d^3}\Big]\exp{(ik\beta d)}\\&+\Big[-i\frac{\Phi(z^-,2,1-\beta)}{2\pi k^2d^2}+\frac{\Phi(z^-,3,1-\beta)}{2\pi k^3d^3}\Big]z^-\exp{(-ik\beta d)}
\end{split}
\end{equation}
and
\begin{equation}\label{a21Leq}
\begin{split}
&a_{21}(k_x)=\Big[-i\frac{\Phi(z^+,2,1-\beta)}{2\pi k^2d^2}+\frac{\Phi(z^+,3,1-\beta)}{2\pi k^3d^3}\Big]\\&\times z^+\exp{(-ik\beta d)}+\Big[-i\frac{\Phi(z^-,2,\beta)}{2\pi k^2d^2}+\frac{\Phi(z^-,3,\beta)}{2\pi k^3d^3}\Big]\\&\times\exp{(ik\beta d)},
\end{split}
\end{equation}
where $\Phi(z,s,a)$ is the Lerch transcendent that is given as $\Phi(z,s,a)=\sum_{n=0}^\infty z^n/(n+a)^s$ \cite{NISThandbook}. 

In fact, Eq.(\ref{eigenvalue_long}) corresponds to an eigenvalue problem that gives rises to the dispersion relation (or band structure) of the longitudinal eigenmodes, and the matrix in the LHS can be regarded as the effective Hamiltonian $H(k_x)$ in the reciprocal space \cite{pocockArxiv2017}. As a result, the eigenvalue problem results in a two-band dispersion relation as follows
\begin{equation}\label{dispersionrelation}
\begin{split}
&\frac{1}{4\pi (ka)^3}\frac{\varepsilon_p(\omega)+2}{\varepsilon_p(\omega)-1}-\frac{i}{6\pi }=a_{11}(k_x)\\&\pm\sqrt{a_{12}(k_x)}\sqrt{a_{21}(k_x)},
\end{split}
\end{equation} 
from which the eigenfrequencies can be solved in the lower complex plane with respect to a fixed $k_x$ and the band structure is then obtained straightforwardly. The solved eigenfrequencies are generally expressed in the complex form of $\tilde{\omega}=\omega-i\Gamma/2$, where the real part $\omega$ stands for the angular frequency of the eigenmode while the imaginary part $\Gamma$ corresponds to its linewidth (or decay rate of the eigenmode) \cite{caoRMP2015,pocockArxiv2017,wang2018topological,wangPRB2018b}. Actually, if a lossless perfect metallic material with a real permittivity in the form of Eq.(\ref{permittivity}) with $\gamma=0$ is considered, by using the quasistatic approximation (namely, no retardation effect) and taking only the nearest-neighbor coupling into account, like the case studied by Ling \textit{et al.} \cite{lingOE2015}, we can immediately find that $a_{11}(k_x)=0$, and $a_{12}(k_x)$ and $a_{21}(k_x)$ are both real. This treatment gives rise to an Hermitian eigenvalue problem and thus a purely real photonic band structure \cite{lingOE2015}. However, it should be noted that this is an ideal case which is valid only when $kd\ll1$ and material dissipation is neglected, and hence is not a general situation encountered in our system, where $kd$ may approach or even be larger than 1 and dissipation as well as retardation effect is substantial \cite{zhangPRB2018}. These effects make the present system become non-Hermitian. Therefore, we must study the topological properties of the band structures in the form of complex eigenfrequencies, as will shown in detail in Section \ref{effect_of_d_sec} where relatively large lattice constants are encountered.

\subsection{Finite chains}
In fact, the calculation of the band structures of finite chains is rather straightforward and can also be done by using Eq.(\ref{coupled_dipole_eq}) with a zero incident field \cite{weberPRB2004,pocockArxiv2017}. In this way, we obtain an eigenvalue equation in the form of
\begin{equation} \mathbf{G}|\mathbf{p}\rangle=\alpha^{-1}(\omega)|\mathbf{p}\rangle.
\end{equation} 
Here $\mathbf{G}$ stands for the interaction Green's matrix whose elements are derived from the Green's function (Eq.(\ref{Gxx})), and $|\mathbf{p}\rangle=[p_1p_2...p_j...p_N]$ is the right eigenvector, which stands for the dipole moment distribution of an eigenmode, where $p_j$ is the dipole moment of the $j$-th NP. Like the case of infinite chains, this equation also specifies a set of complex eigenfrequencies in the lower complex plane in the form of $\tilde{\omega}=\omega-i\Gamma/2$. The physical significance of these complex eigenfrequencies are the same as the ones appearing in Section \ref{model_infinite}. In addition, to quantify the degree of eigenmode localization, we further calculate the inverse participation ratio (IPR) of an eigenmode from its eigenvector as \cite{wangOL2018,wang2018topological}. 
\begin{equation}
\mathrm{IPR}=\frac{\sum_{n=1}^{N}|p_j|^4}{(\sum_{n=1}^{N}|p_j|^2)^2}.
\end{equation}
For an IPR approaches $1/M$, where $M$ is an integer, the corresponding eigenmode involves the excitation of $M$ NPs \cite{wangOL2018}. For a highly localized topological eigenmode, its IPR should be much larger than those of the bulk eigenmodes \cite{wangOL2018}. Therefore, this quantity provides an indicator for the topological edge modes.



\section{Results and Discussion}
In this paper, we mainly investigate the topological properties of longitudinal modes. This is because transverse ones are more strongly coupled to the free-space radiation with a much narrower band gap and the localization degree is lower due to the long-range dipole-dipole interactions. These features make it difficult to observe transverse topological eigenmodes experimentally, as discussed in our previous papers \cite{wang2018topological,wangPRB2018b}. Here we first calculate the Bloch band structures, from which the complex Zak phase can be determined by using the biorthogonality of the Bloch eigenvectors in order to characterize the topological properties of the system. Secondly, we study a finite lattice and numerically calculate the eigenmode distribution, in which the topologically protected edge modes, namely, TPPs, are demonstrated. Thirdly, the effects of lattice constant on the eigenmode distribution and edge modes are investigated. Finally, we discuss the effects of doping type and concentration on the properties of TPPs as a demonstration for their wideband tunability.

\subsection{Bloch band structures and the complex Zak phase}\label{bulkside_sec}
\begin{figure}[htbp]
	\centering
	\subfloat{
		\includegraphics[width=0.46\linewidth]{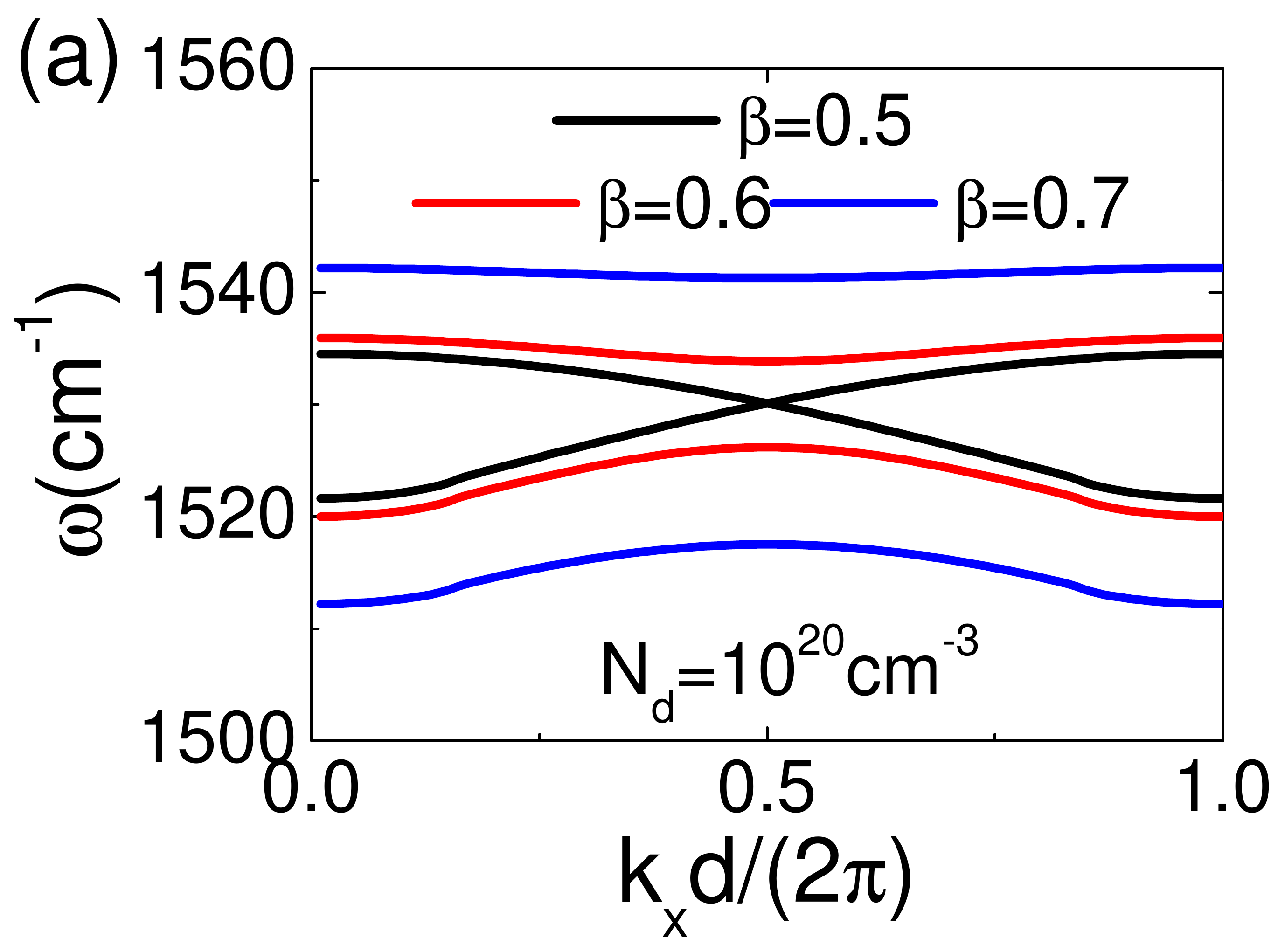}\label{bandstructurelongd1}
	}
	\hspace{0.01in}
	\subfloat{
		\includegraphics[width=0.46\linewidth]{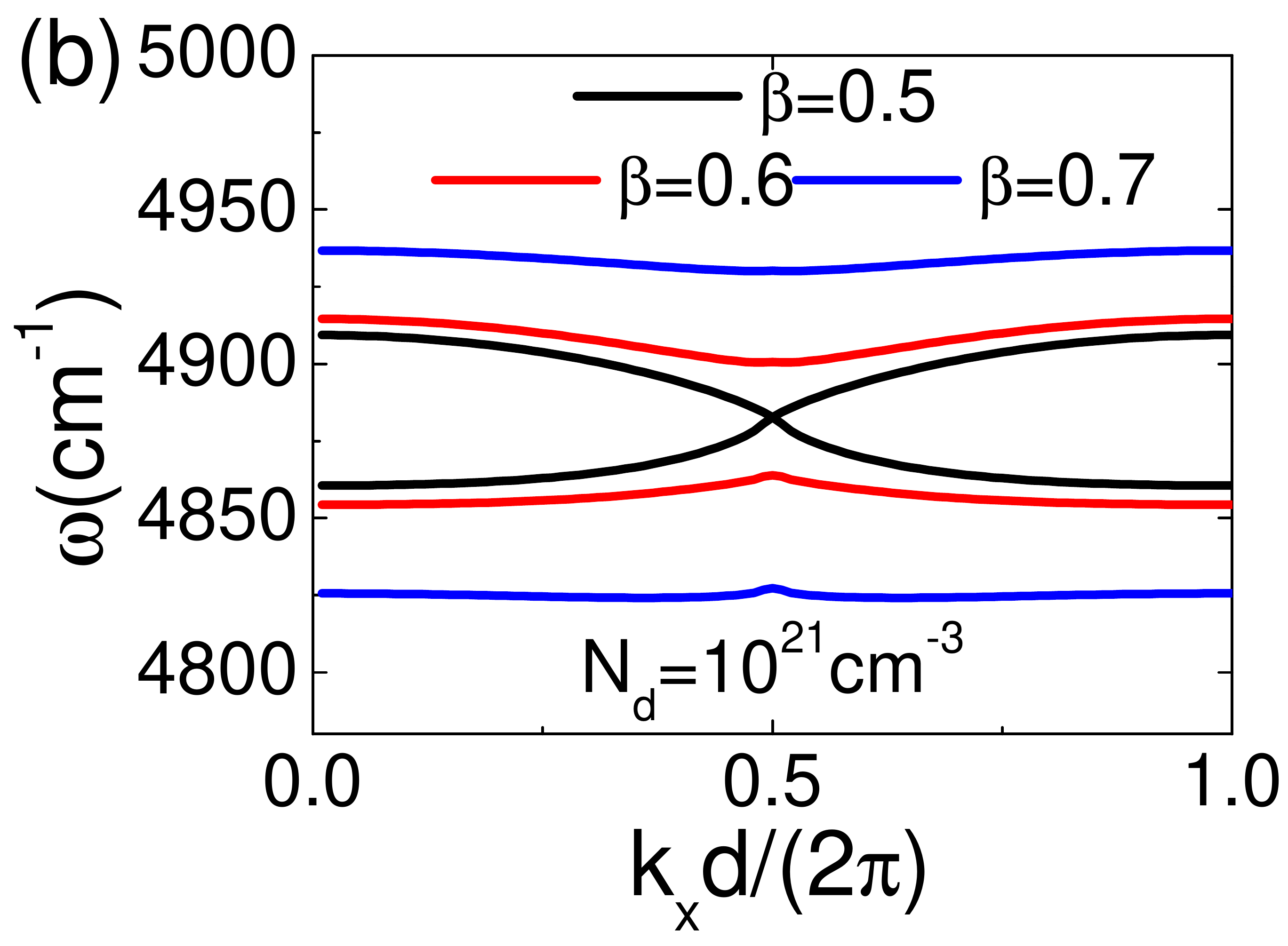}\label{bandstructurelongd1N21}
	}
	\caption{Real parts of the longitudinal band structures of a dimerized $n$-type doped Si NP chain with a doping level of (a) $N_d=1\times10^{20}\mathrm{cm}^{-3}$ and (b) $N_d=1\times10^{21}\mathrm{cm}^{-3}$. The lattice constant is $d=d_1+d_2=1\mathrm{\mu m}$ with different dimerization parameters $\beta$. Note the Bloch band structures are the same for $\beta$ and $1-\beta$ when other system parameters are fixed.}\label{figcda}
	
\end{figure} 

We first investigate the band structure of a dimerized chain composed of $n$-doped Si NPs with an overall lattice constant of $d=1\mathrm{\mu m}$. The doping concentration is set to be $N_d=1\times10^{20}\mathrm{cm}^{-3}$. In Figs.\ref{bandstructurelongd1} and \ref{bandstructurelongd1N21} we show the real parts of the bulk band structures under two different doping levels of $N_d=1\times10^{20}\mathrm{cm}^{-3}$ and $N_d=1\times10^{21}\mathrm{cm}^{-3}$, for different dimerization parameters $\beta=0.5, 0.6, 0.7$ for both longitudinal eigenmodes respectively. Note the bulk band structures for $\beta=0.3$ and $\beta=0.4$ are the same as those for $\beta=0.7$ and $\beta=0.6$ correspondingly. In general, the band structure is the same for the cases of $\beta$ and $1-\beta$, while the difference of these two cases lies in their topological invariant \cite{pocockArxiv2017,downingPRB2017}. It is also found that for $\beta\neq0.5$, band gaps in the real frequency space are opened in both cases and a larger $|\beta-0.5|$ (i.e., the deviation from a mono-atomic chain) gives rise to a wider band gap. This behavior is consistent with the conventional SSH model \cite{atalaNaturephys2013}. Moreover, for different doping concentrations, the central frequency of the band gap is very close to the LSPR mode of a single NP, because the band gap is opened, in essence, due to the strong coupling between the collective SPP modes of the two constituting sublattices for $\beta\neq0.5$ \cite{wangPRB2018b}. On the other hand, the imaginary spectrum is ungapped for all dimerization parameters (not shown here) \cite{wangPRB2018b}.

In 1D non-Hermitian systems, the complex Zak phase, as the geometric phase picked up by an eigenmode when it adiabatically evolves across the first Brillouin zone (BZ), can be conveniently used to determine the topology of bulk band structure, as indicated by Lieu \cite{lieuPRB2018}. It is determined by the integration of the Berry connection over BZ as \cite{yucePLA2015,yucePRA2018,lieuPRB2018,alvarezPRB2018,wang2018topological,wangPRB2018b}  
\begin{equation}\label{cberryphase}
\begin{split}
\theta_\mathrm{Z}&=\int_\mathrm{BZ}dk_x\mathcal{A}(k_x),
\end{split}
\end{equation}
where $\mathcal{A}(k_x)$ is the Berry connection that can be determined from the normalized left and right eigenvectors $|p_{k_x}^{L}\rangle$ and $|p_{k_x}^{R}\rangle$ ($\langle p_{k_x}^L|p_{k_x}^R\rangle=1$) for longitudinal eigenmodes as
\begin{equation}\label{cberryphase2}
\begin{split}
\mathcal{A}(k_x)=i\Big[p_{A,k_x}^{L,*}\frac{\partial p_{A,k_x}^R}{\partial k_x}+p_{B,k_x}^{L,*}\frac{\partial p_{B,k_x}^R}{\partial k_x}\Big].
\end{split}
\end{equation}
The left and right eigenvectors are calculated as
\begin{equation}\label{lefteigenvector}
|p_{k_x}^{L}\rangle=\left(\begin{matrix}p_{A,k_x}^{L}\\p_{B,k_x}^{L}\end{matrix}\right)=\frac{1}{\sqrt{2}}\left(\begin{matrix}\mp\frac{\sqrt{a_{21}^{*}(k_x)}}{\sqrt{a_{12}^{*}(k_x)}}\\1\end{matrix}\right),
\end{equation}
\begin{equation}\label{righteigenvector}
|p_{k_x}^{R}\rangle=\left(\begin{matrix}p_{A,k_x}^{R}\\p_{B,k_x}^{R}\end{matrix}\right)=\frac{1}{\sqrt{2}}\left(\begin{matrix}\mp\frac{\sqrt{a_{12}(k_x)}}{\sqrt{a_{21}(k_x)}}\\1\end{matrix}\right).
\end{equation}
Note the left eigenvector is solved through the relation of  $H^\dag(k_x)|p_{k_x}^{L}\rangle=E_{k_x}^*|p_{k_x}^{L}\rangle$.
Therefore, the complex Zak phase is given by
\begin{equation}\label{cberryphase3}
\begin{split}
\theta_\mathrm{Z}=\frac{\arg[a_{21}(k_x)]-\arg[a_{12}(k_x)]}{4}\Big|_{-\pi/d}^{\pi/d},
\end{split}
\end{equation}
which is related to the difference of the winding of the off-diagonal elements of the effective Hamiltonian around the origin. It is shown that although the present system is non-Hermitian and exhibits a breaking of chiral symmetry, the complex Zak phase is still quantized (only two values, 0 and $\pi$, are allowed) in a similar way as in a chirally symmetric system  due to the fact that the eigenvectors are independent of the diagonal elements of the effective Hamiltonian \cite{pocockArxiv2017,wang2018topological,wangPRB2018b}. 
According to Eq.(\ref{cberryphase3}), the complex Zak phase is actually a real quantity \cite{wang2018topological}, and it is simply half the difference of the winding numbers of $a_{21}(k_x)$ and $a_{12}(k_x)$ encircling the origin multiplied by $\pi$.  Since the directions of the encircling of $a_{12}(k_x)$ and $a_{21}(k_x)$ are always opposite because $a_{12}(k_x)=a_{21}(-k_x)$, the winding numbers of $a_{12}(k_x)$ and $a_{21}(k_x)$ are +1 and -1 respectively when the dimerization parameter is $\beta=0.7$ and $\beta=0.6$, and are both zero when $\beta=0.3$ and $\beta=0.4$. Therefore, the complex Zak phase for $\beta=0.7$ and $\beta=0.6$ is $\pi$ and is $0$ for $\beta=0.3$ and $\beta=0.4$. Furthermore, we have examined that irrespective of the lattice constant and doping concentration, the complex Zak phase is 0 for $\beta<0.5$ and $\pi$ for $\beta>0.5$ for longitudinal modes \cite{wang2018topological,wangPRB2018b}. As a result, this complex Zak phase is a well-defined topological invariant in the bulk side. In the next subsection, we will check the bulk-boundary correspondence by identifying topological edge modes of a finite system.

\subsection{Bulk-boundary correspondence and midgap modes}
Recently, there has been growing attention on the topological properties of non-Hermitian systems. In Hermitian systems, the principle of bulk-boundary correspondence indicates that the topological invariant determines the existence of edge modes, and the total winding number $\mathcal{W}=\theta_\mathrm{Z}/\pi$ is equivalent to the number of edge modes localized over the boundary of the systems \cite{luNPhoton2014,khanikaevNPhoton2017,ozawa2018topological}. However, it was surprisingly found that the conventional bulk-boundary correspondence becomes invalid for some specific 1D non-Hermitian Hamiltonians \cite{yao2018edge,kunst2018biorthogonal,xiongJPC2018}. This is because in those systems, the Bloch band structures under the periodic boundary condition are much different from the band structures calculated from the open boundary condition \cite{yao2018edge,kunst2018biorthogonal}. Therefore, to appropriately investigate the topological properties of the present non-Hermitian system, it is necessary to directly compute the band structures of finite chains with open boundaries.

\begin{figure}[htbp]
	\centering
	\subfloat{
	\includegraphics[width=0.46\linewidth]{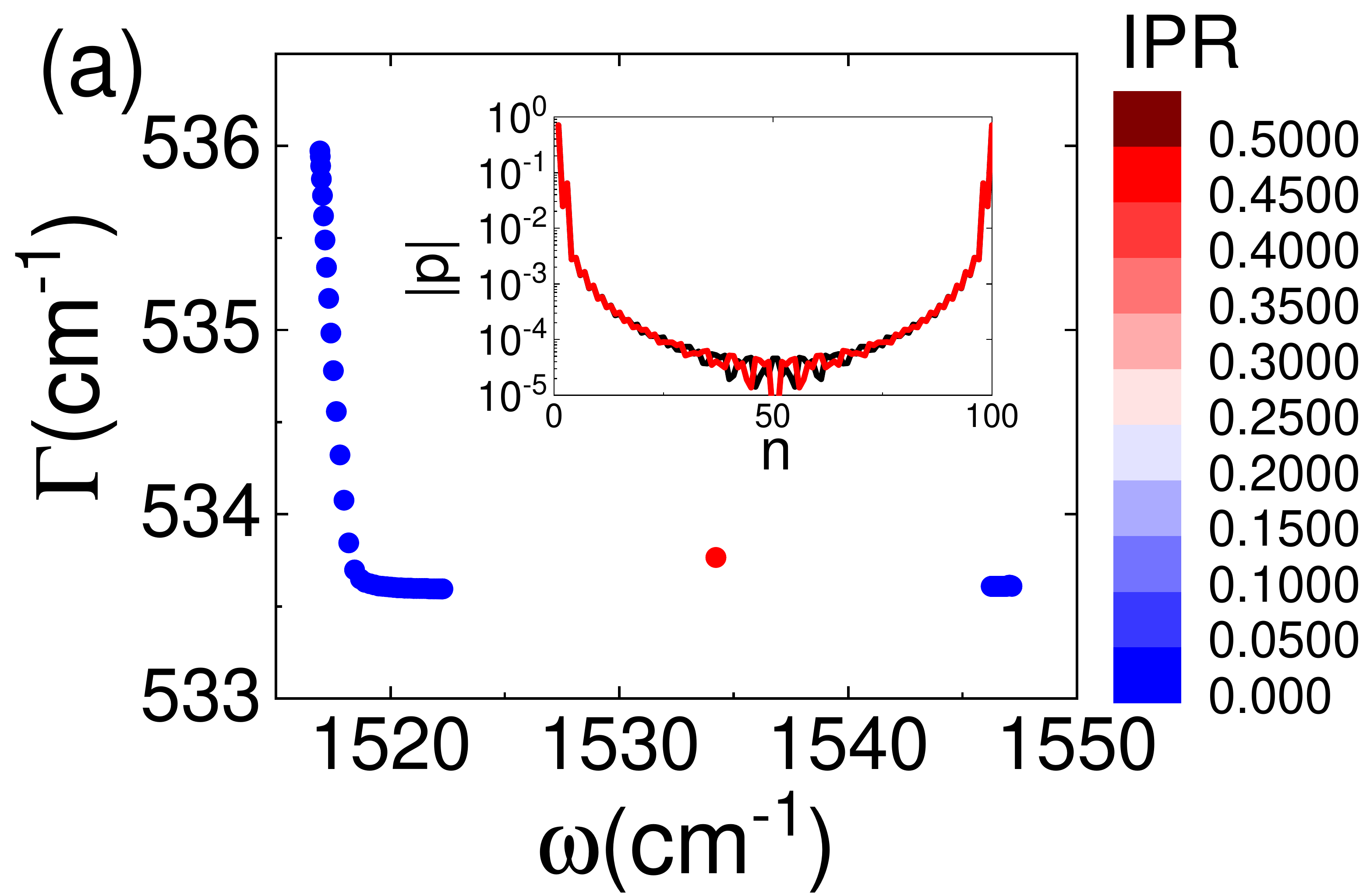}\label{beta07longband}
}
	\hspace{0.01in}
	\subfloat{
	\includegraphics[width=0.46\linewidth]{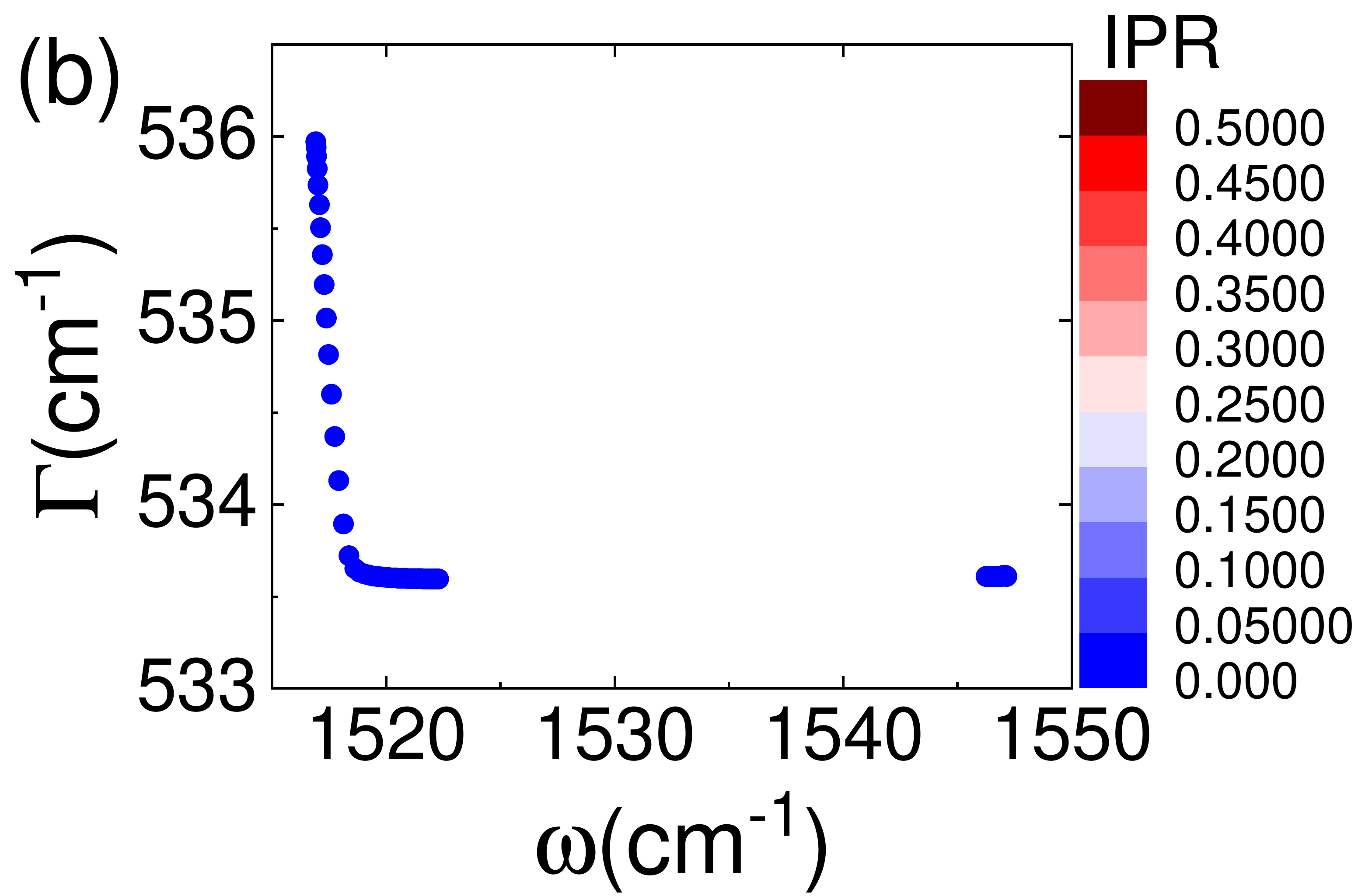}\label{beta03longband}
}


	\caption{(a) Longitudinal eigenmode distribution of a dimerized chain with $N=100$ NPs under $\beta=0.7$ and $d=1\mathrm{\mu m}$. Note there are two midgap modes. (b) The same as (a) but here $\beta=0.3$. Inset: Dipole moment distribution of the midgap edge modes.} \label{eigenmodelong}
\end{figure}

In Figs.\ref{beta07longband} and \ref{beta03longband}, we show the eigenmode distribution of the longitudinal modes in the complex frequency plane for $\beta=0.7$ and $\beta=0.3$ with a lattice constant of $d=1\mathrm{\mu m}$. The NPs are $n$-doped with a doping concentration of $N_d=1\times10^{20}\mathrm{cm}^{-3}$. The number of NPs in the chain is fixed as 100 (50 per sublattice). In both cases we can find that band gaps are opened in the complex frequency plane. The corresponding frequency range of the band gaps are also the same as the Bloch band structure. The only difference between the $\beta=0.7$ and $\beta=0.3$ cases is that there are two midgap modes with high IPRs in the band gap in the former case. The dipole moment distributions of the two midgap modes are shown in the inset of Fig.\ref{beta07longband}, which show a highly localized shape from the boundary. Therefore, by noting the nontrivial complex Zak phase of the $\beta=0.7$ case in the bulk side, we can conclude that these midgap modes are topologically protected edge modes.

\begin{figure}[htbp]
	\centering
	\subfloat{
		\includegraphics[width=0.5\linewidth]{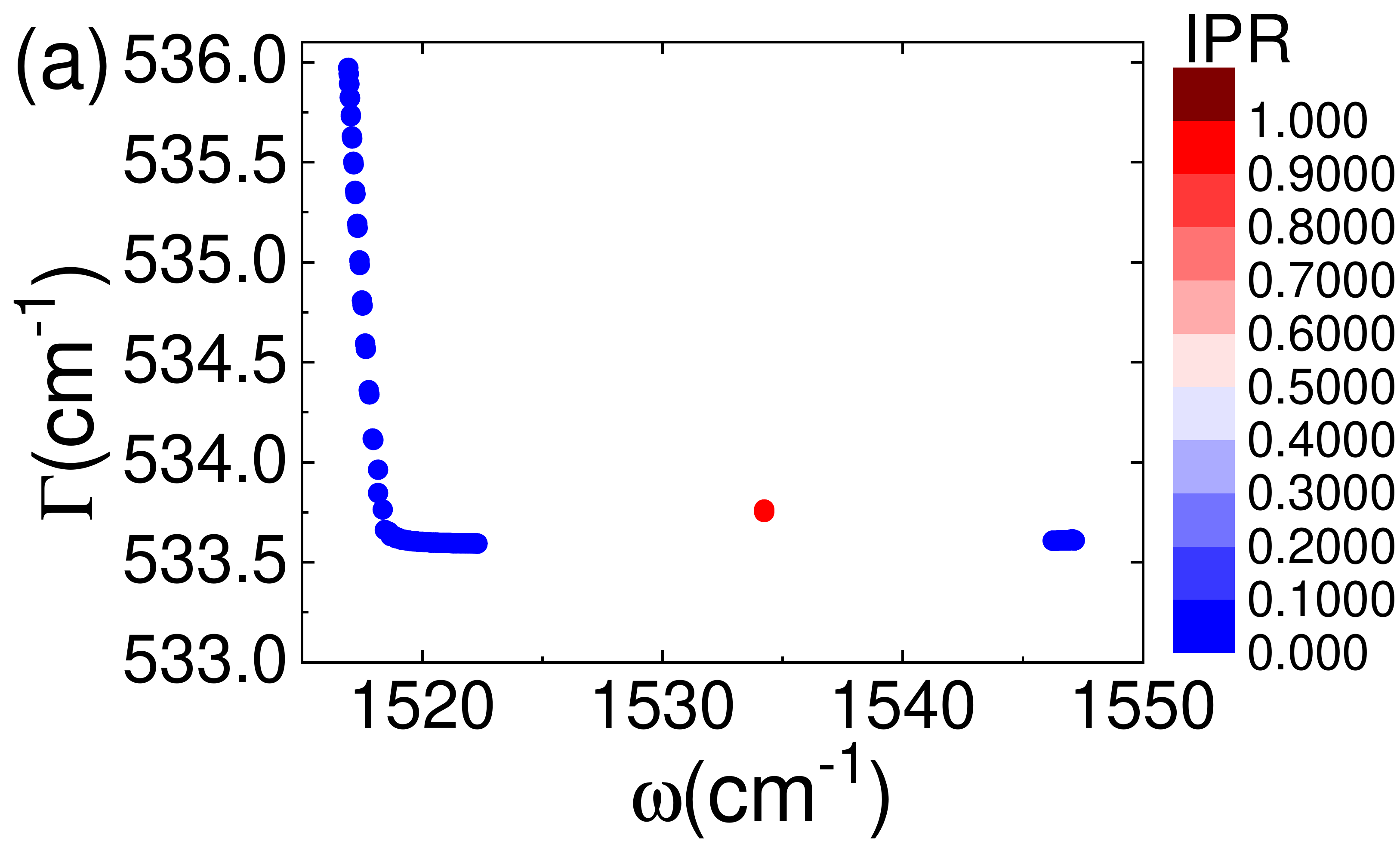}\label{interfacemodebeta07}
	}
	\hspace{0.01in}
	\centering
	\subfloat{
		\includegraphics[width=0.43\linewidth]{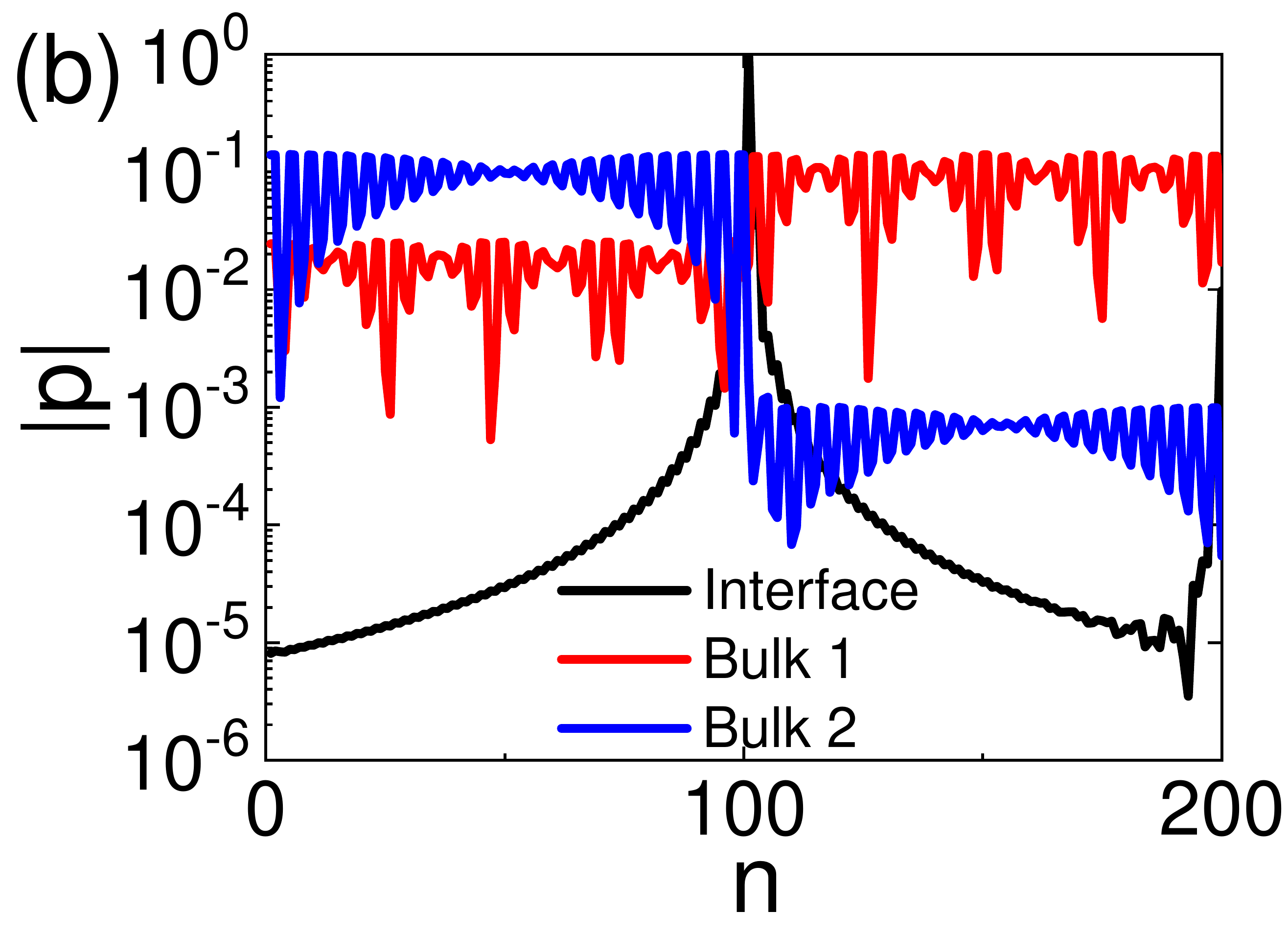}\label{interfacemodedipole}
	}
	
	\caption{(a) Longitudinal eigenmode distribution for a connected chain. (b) Dipole moment distribution of the interface  mode at ($\omega=1534.24\mathrm{cm}^{-1}$, $\Gamma=533.75\mathrm{cm}^{-1}$) in (d), compared with those of bulk eigenmodes at ($\omega=1519.84\mathrm{cm}^{-1}$, $\Gamma=533.60\mathrm{cm}^{-1}$) and ($\omega=1546.70\mathrm{cm}^{-1}$, $\Gamma=533.61\mathrm{cm}^{-1}$).} \label{connectedchain}
\end{figure}

To further examine the bulk-boundary correspondence, we can check whether the topologically protected interface states can emerge at the boundary of two topologically different chains. In Fig.\ref{interfacemodebeta07}, the eigenmode distribution of a 1D connected chain is presented, which consists of a topologically trivial chain with $\beta=0.3$ in the left and a topologically nontrivial chain with $\beta=0.7$ in the right. The distance between the two chains is set to be $1\mathrm{\mu m}$ (the distance between the centers of the rightmost NP in the left chain and the leftmost NP in the right chain.). Two midgap modes with high IPRs are also observed, in which one is the interface mode while the other is the edge mode localized at the right boundary of the right chain. Their IPRs are much larger than the midgap modes in the single chain case, both larger than 0.9. In Fig.\ref{interfacemodedipole} we show the dipole moment distribution for the interface mode, which is strongly localized over the interface. And the dipole moments of two typical bulk eigenmodes are also given for comparison. Therefore, we can unequivocally conclude that the complex Zak phase defined in Section \ref{bulkside_sec} is able to characterize these longitudinal edge modes, i.e., TPPs, and the principle of bulk-boundary correspondence is valid in this circumstance despite the non-Hermiticity.

In a word, from the eigenmode distributions of finite chains, highly localized TPPs can be found, which are topologically protected by the complex Zak phase if the dimerization parameter $\beta>0.5$, implying the validity of the bulk-boundary correspondence. This behavior is the same as that of the conventional SSH model, since the short-range near-field dipole-dipole interactions for the longitudinal modes dominate in this situation \cite{wangPRB2018b} and hence the non-Hermiticity is weak \cite{wang2018topological}.

\subsection{Effect of lattice constant}\label{effect_of_d_sec}
\begin{figure}[htbp]
	\centering
	\subfloat{
		\includegraphics[width=0.46\linewidth]{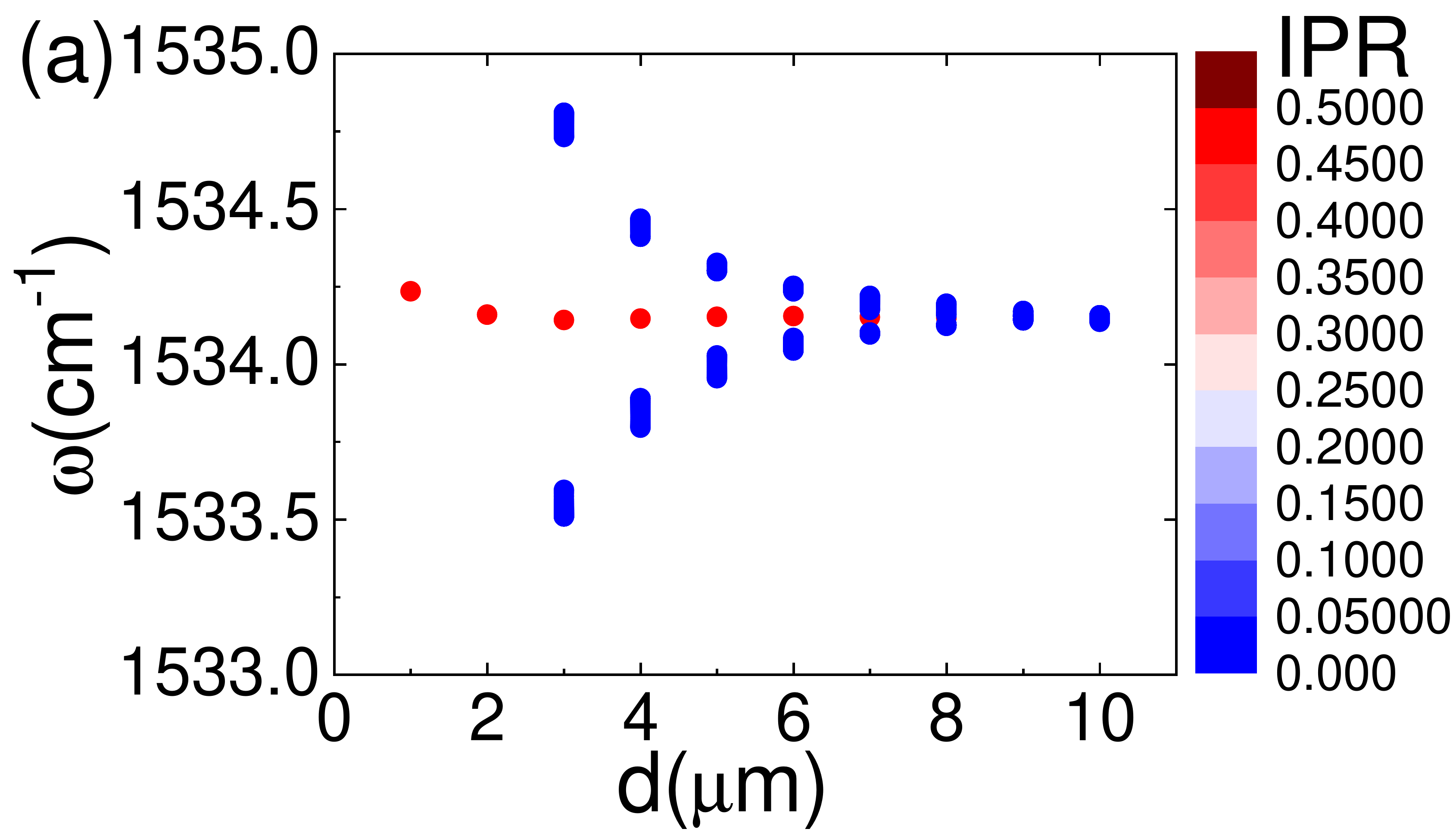}\label{bandevolutionlongreal}
	}
	\hspace{0.01in}
	\subfloat{
		\includegraphics[width=0.46\linewidth]{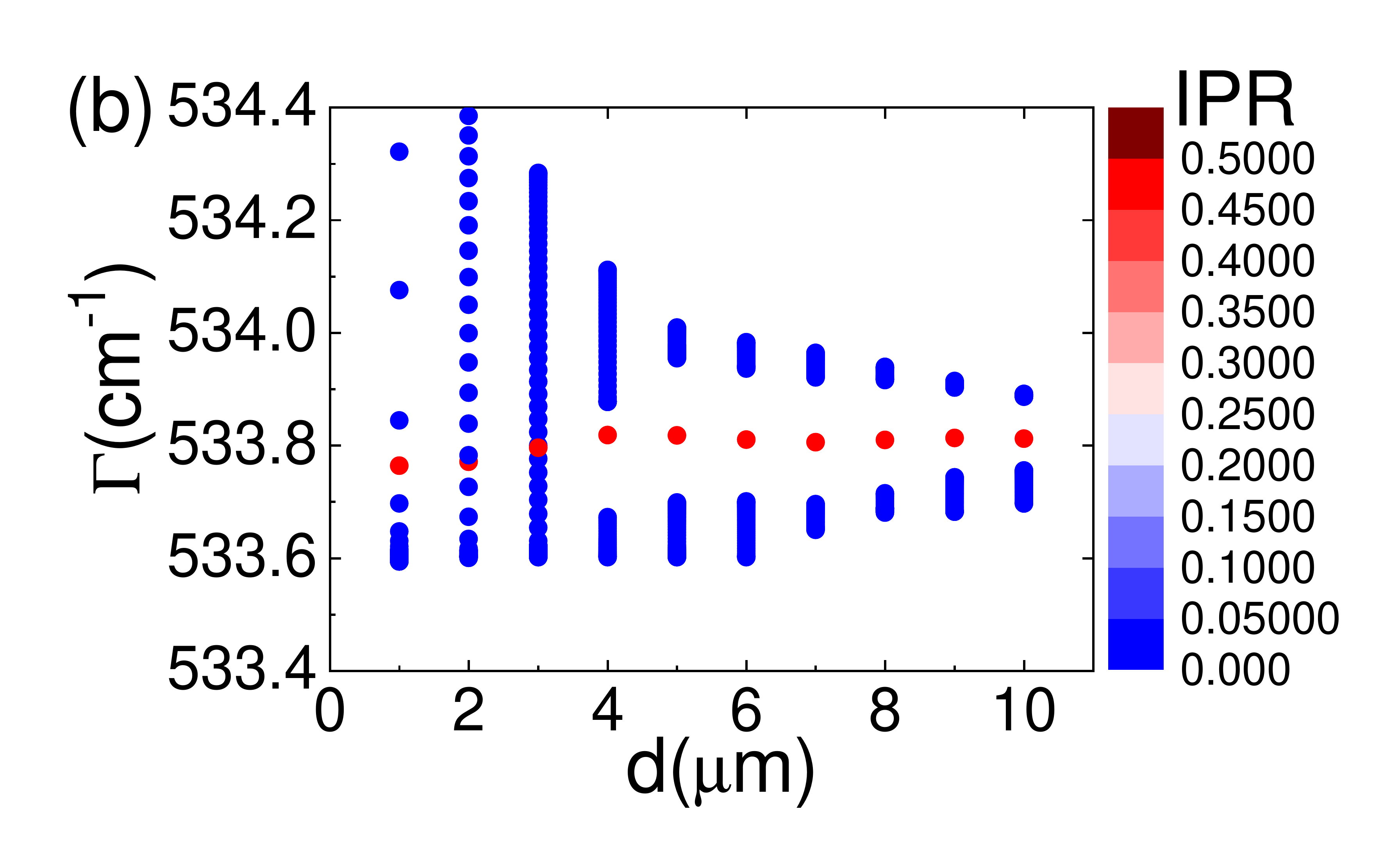}\label{bandevolutionlongimag}
	}
	\hspace{0.01in}
	\subfloat{
		\includegraphics[width=\linewidth]{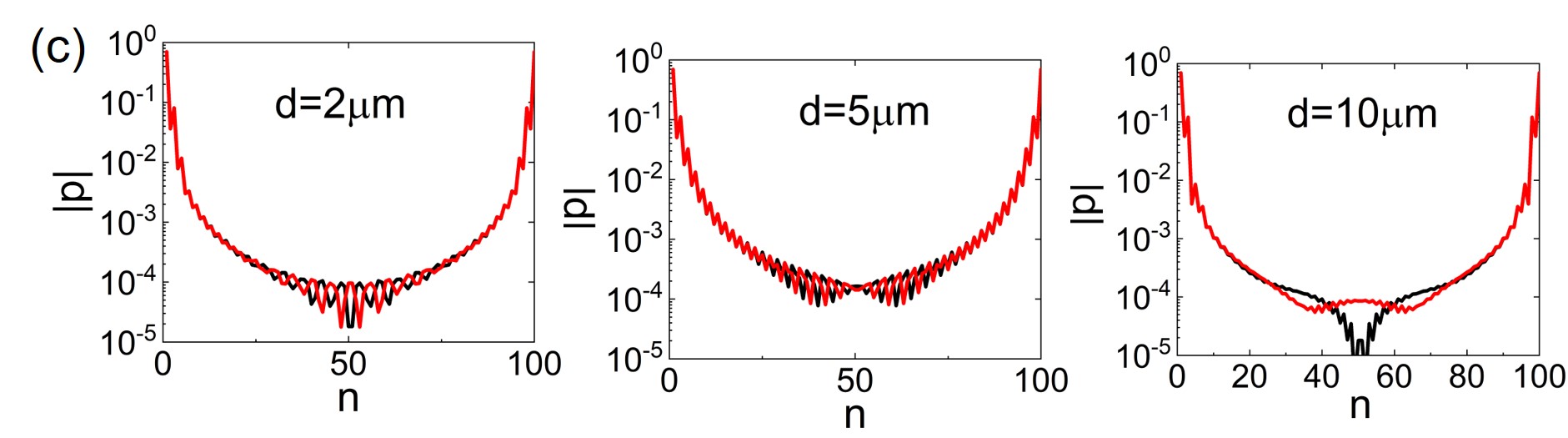}\label{midgapmodewithd}
	}
	\caption{Real (a) and imaginary (b) parts of the complex eigenfrequency spectrum of longitudinal eigenmodes as a function of the lattice constant $d$ for $\beta=0.7$. (c) Dipole moment distribution for the topological edge modes at different lattice constants.}\label{bandedgeevolutionlong}
\end{figure}

We briefly discuss the effect of lattice constant on the topological edge modes for the longitudinal polarization. In Figs.\ref{bandevolutionlongreal} and \ref{bandevolutionlongimag}, the real and imaginary parts of the eigenfrequency spectrum of a finite chain with 100 NPs as a function of the lattice constant are provided. The NPs are $n$-doped with a doping concentration of $N_d=1\times10^{20}\mathrm{cm}^{-3}$. The dimerization parameter is fixed as $\beta=0.7$. It is seen that the complex band gaps are always open and eigenmodes with high IPRs are well situated in the complex band gaps, despite the fact the band gaps in the real-frequency plane are almost closed at large lattice constants. More specifically, for $d\gtrsim6 \mathrm{\mu m}$, the real part of the band gap nearly closes while we observe the opening of the imaginary part of the band gap, and hence the high-IPR modes are still midgap modes. In addition, the complex frequency of the topological modes does not vary much with the increase of the lattice constant. The dipole moment distributions of the topological midgap modes are presented in Fig.\ref{midgapmodewithd} for three different lattice constants $d=2,~5$ and $10\mathrm{\mu m}$, in which all edge modes exhibit a highly localized profile. As a consequence, taking the entire complex band structure into account in a non-Hermitian system is necessary and critical to correctly understanding its topological properties \cite{shenPRL2018,pocockArxiv2017,wang2018topological}. In addition, we also confirm that for the cases of $\beta<0.5$, no localized edge eigenmodes are observed. The bulk-boundary correspondence for longitudinal modes is thus valid at different lattice constants as predicted by the quantized complex Zak phase in Section \ref{bulkside_sec}. Therefore, in a word, for different lattice constants, we are able to find topological plasmon polaritons in the present system.

\subsection{Doping dependence of TPPs}
\begin{figure}[htbp]
	\centering
	\subfloat{
		\includegraphics[width=0.46\linewidth]{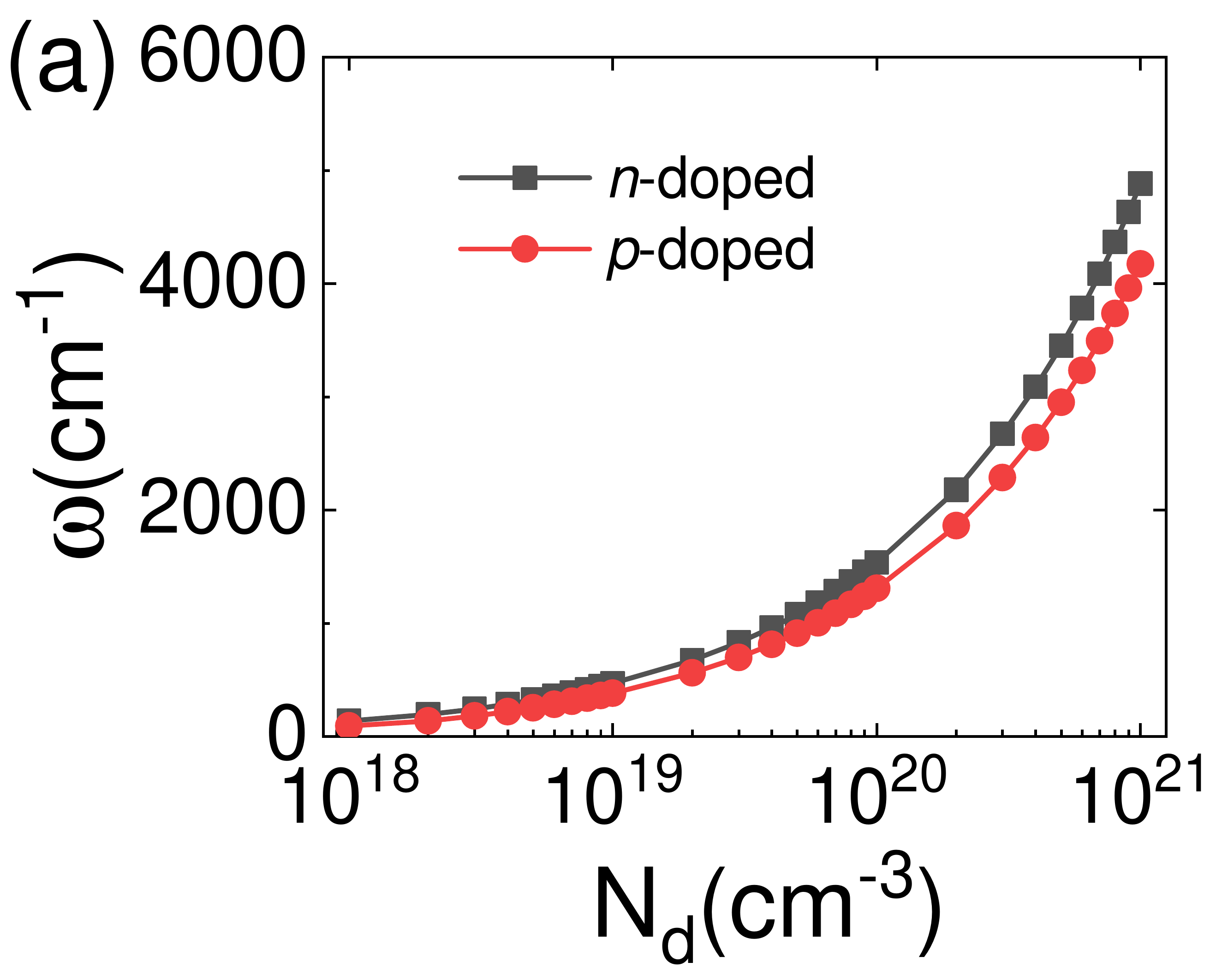}\label{dopingeffect}
	}
    \hspace{0.01in}
	\subfloat{
	\includegraphics[width=0.46\linewidth]{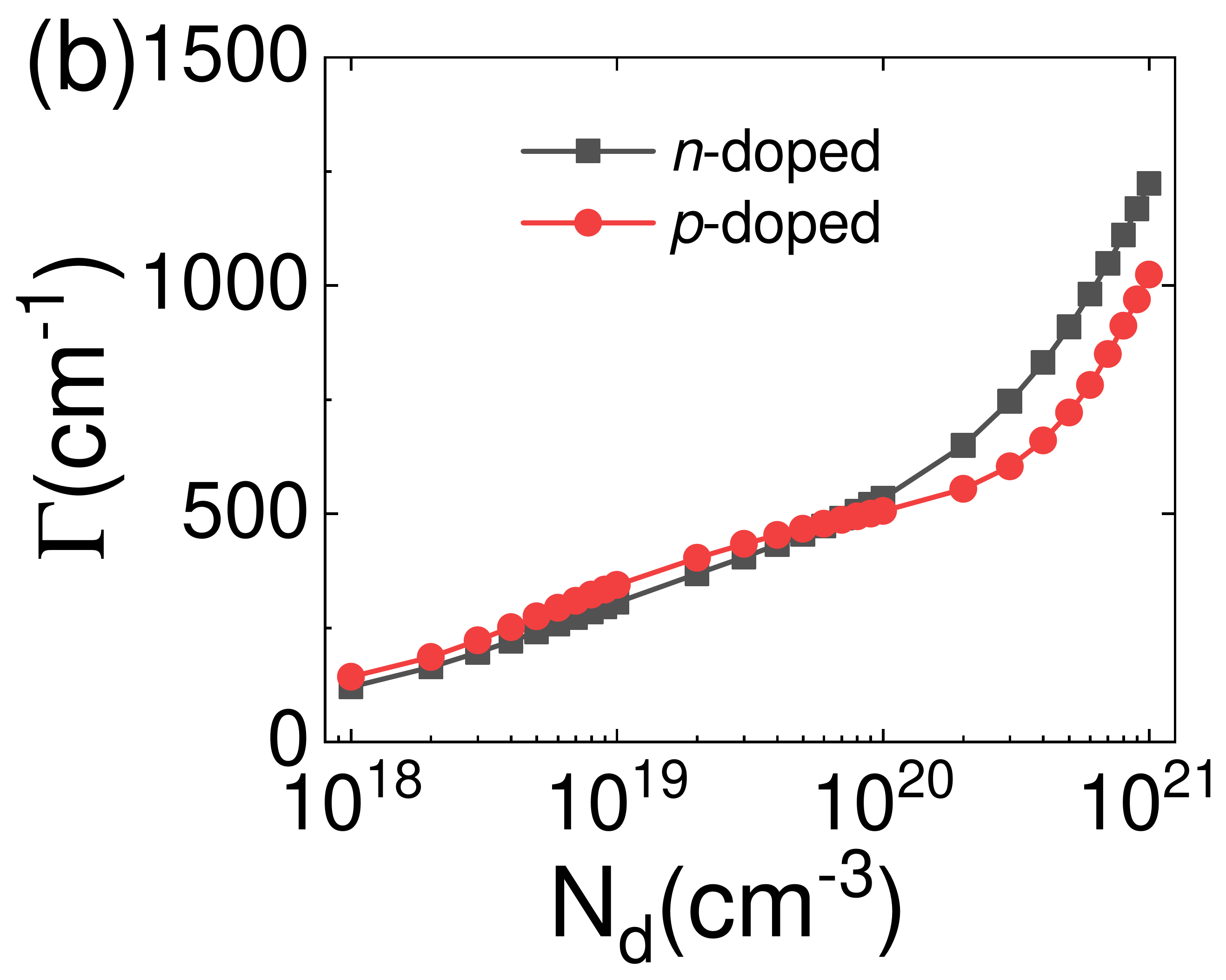}\label{dopingeffectgamma}
}

	\caption{Effects of doping type and concentration on the (a) resonance frequency, (b) decay rate of the topological plasmon polaritons for $\beta=0.7$ and $d=1 \mathrm{\mu m}$.}\label{doping1}
	
\end{figure}

In Fig.\ref{doping1}, we show the variation of resonance frequency and decay rate of TPPs as a function of the doping concentration for both $n$-type and $p$-type Si NP chains. Note for a high doping concentration, the resonance wavelength becomes small, which might lead the dipole approximation to be invalid. We have verified that in the investigated range of doping concentration up to $N_d=1\times10^{21}~\mathrm{cm}^{-3}$, the dipole approximation remains to be valid. In the studied parameter range, the complex Zak phase is well-defined and the bulk-boundary correspondence is valid, leading to well-defined TPPs in the complex band gaps for $\beta>0.5$.

It is found in Fig.\ref{doping1} that the resonance frequency of TPPs can be varied from far-infrared (at a angular frequency around 100 cm$^{-1}$) to near-infrared (at a angular frequency about 4800 cm$^{-1}$), and for a fixed doping concentration, the resonance frequency of TPPs in $n$-type Si NP chains is slightly higher than that in $p$-type Si NP chains. On the other hand, the decay rate of TPPs in $p$-type Si NP chains is slightly larger than that in $n$-type Si NP chains when the doping concentration is lower than $\sim8\times10^{19}\mathrm{cm}^{-3}$. Further increasing the doping concentration leads the decay rate of TPPs in $n$-type Si NP chains to become larger instead.

\begin{figure}[htbp]
	\centering
	\subfloat{
		\includegraphics[width=0.46\linewidth]{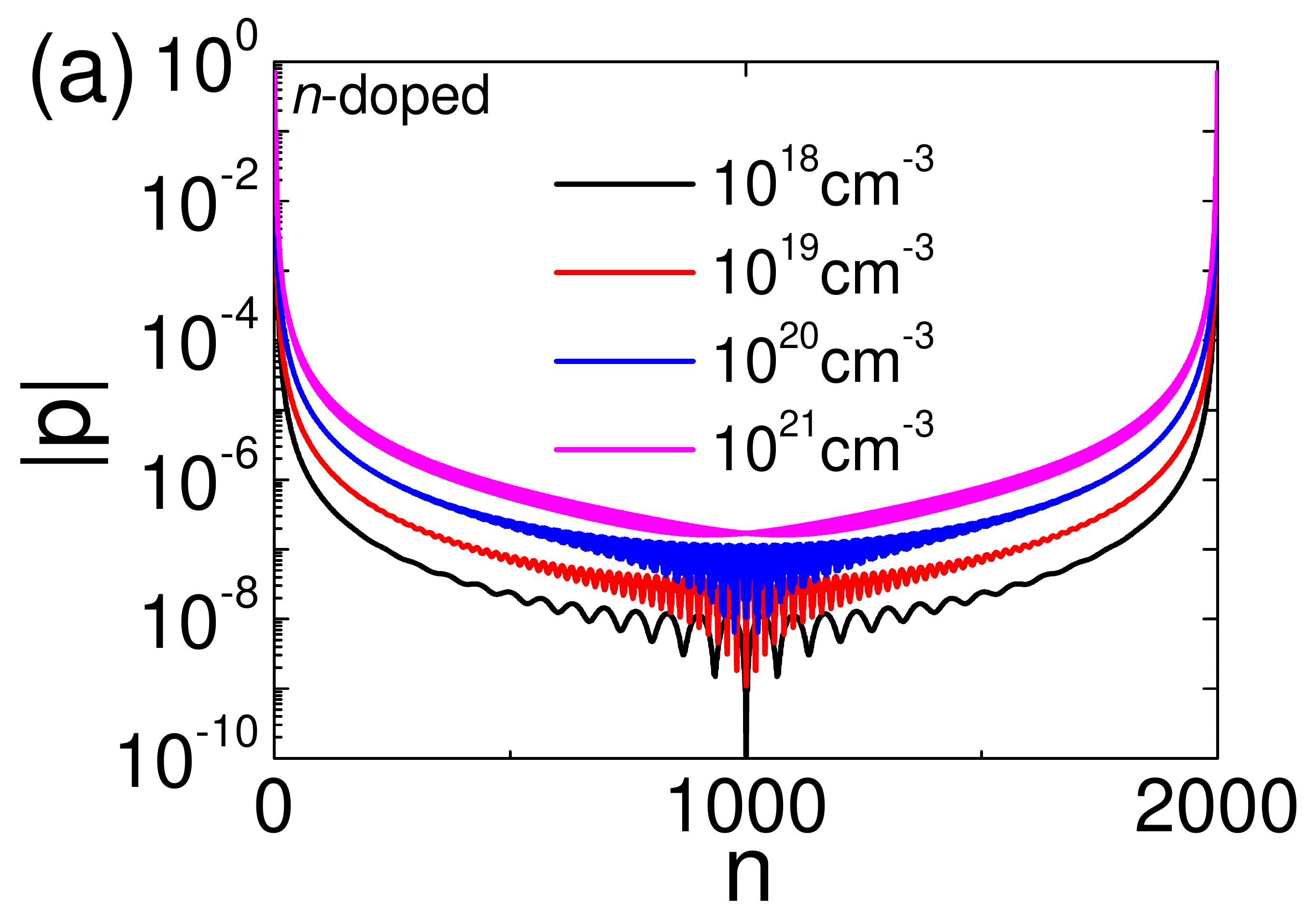}\label{midgapmodendopeN2000}
	}
	\subfloat{
		\includegraphics[width=0.46\linewidth]{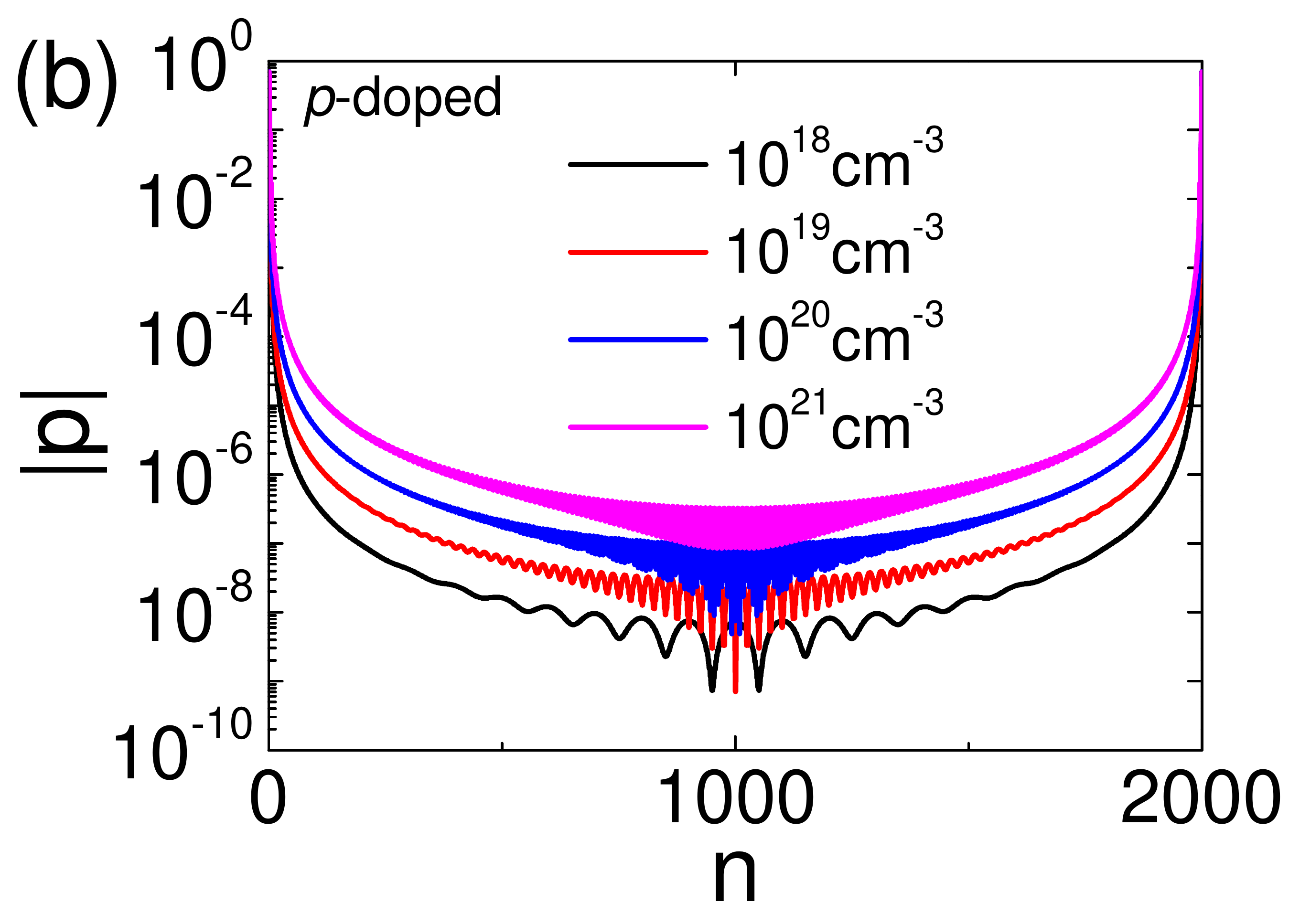}\label{midgapmodepdopeN2000}
	}
	
	\caption{Effects of doping type and concentration on the dipole moment distribution of the topological plasmon polaritons for $\beta=0.7$ and $d=1 \mathrm{\mu m}$. Here $N=2000$ NPs are used. (a) $n$-type. (b) $p$-type. }\label{doping2}
	
\end{figure} 

In Fig.\ref{doping2}, the evolution of dipole moment distribution of the topologically protected plasmon polaritons with doping type and concentration is shown at fixed geometric parameters ($d=1\mathrm{\mu m}$ and $\beta=0.7$). To see the localization behavior more clearly, here $N=2000$ is used for the chain. We can find that for higher doping concentrations, the localization length of the topological edge modes become longer, which is a result of smaller interparticle dipole-dipole interactions \cite{wangPRB2018b}. To be more specific, with higher concentrations, the frequency of the topological modes becomes higher, leading to a shorter resonance wavelength. Then for a fixed lattice constant $d=1\mathrm{\mu m}$, the interparticle distance is more comparable with the wavelength, giving rise to weaker dipole-dipole interactions and hence larger localization length, as indicated by our previous works \cite{wangPRB2018b,wang2018topological}.

\section{Conclusion}
In conclusion, we investigate topologically protected plasmon polaritons in 1D dimerized doped Si NP chains, which mimic the celebrated SSH model. We carry out this study beyond the nearest-neighbor approximation by taking all near-field and far-field dipole-dipole interactions into account. For longitudinal modes, despite the consequences of non-Hermiticity and the breaking of chiral symmetry brought by this treatment, we show that such dimerized chains can still support topological protected midgap modes, i.e., TPPs,  We reveal that in this system, the band topology can be characterized by a quantized complex Zak phase, which indicates a topological phase transition point of $\beta=0.5$. By analyzing the eigenmodes of a finite chain as well as their IPRs, we find topologically protected midgap modes and unequivocally verify the principle of bulk-boundary correspondence. Moreover, by changing the doping type and doping concentration, it is possible to tune the frequency of these topological plasmon polaritons and the localization length of the edge modes are also modulated accordingly. These TPPs offer an efficient tool for robust and enhanced light-matter interactions in the infrared spectrum in a tunable fashion, with potential applications infrared sensing, near-field heat transfer and so on. Our study can also be extended to other electrical tunable semiconductors like gallium arsenide (GaAs) \cite{junNL2013}, indium arsenide (InAs) \cite{liOE2011} and indium tin oxide (ITO) \cite{shirmaneshNL2018}, etc.

\begin{acknowledgments}
We thank the financial support from the National Natural Science Foundation of China (No. 51636004 and No. 51906144), Shanghai Key Fundamental Research Grant (No. 18JC1413300), China Postdoctoral Science Foundation (No. BX20180187 and No. 2019M651493) and the Foundation for Innovative Research Groups of the National Natural Science Foundation of China (No. 51521004).
\end{acknowledgments}

\bibliography{ssh_doped_silicon}
\end{document}